\documentclass[10pt,a4paper,onecolumn]{article}
\usepackage{marginnote}
\usepackage{graphicx}
\usepackage{xcolor}
\usepackage{authblk,etoolbox}
\usepackage{titlesec}
\usepackage{calc}
\usepackage{tikz}
\usepackage{hyperref}
\hypersetup{colorlinks,breaklinks=true,
            urlcolor=[rgb]{0.0, 0.5, 1.0},
            linkcolor=[rgb]{0.0, 0.5, 1.0}}
\usepackage{caption}
\usepackage{tcolorbox}
\usepackage{amssymb,amsmath}
\usepackage{ifxetex,ifluatex}
\usepackage{seqsplit}
\usepackage{xstring}

\usepackage{float}
\let\origfigure\figure
\let\endorigfigure\endfigure
\renewenvironment{figure}[1][2] {
    \expandafter\origfigure\expandafter[H]
} {
    \endorigfigure
}

\usepackage{fixltx2e} % provides \textsubscript
\usepackage[
  backend=biber,
]{biblatex}
\bibliography{paper.bib}

\let\textttOrig=\texttt
\def\texttt#1{\expandafter\textttOrig{\seqsplit{#1}}}
\renewcommand{\seqinsert}{\ifmmode
  \allowbreak
  \else\penalty6000\hspace{0pt plus 0.02em}\fi}

\makeatletter
\let\href@Orig=\href
\def\href@Urllike#1#2{\href@Orig{#1}{\begingroup
    \def\Url@String{#2}\Url@FormatString
    \endgroup}}
\def\href@Notdoi#1#2{\def\tempa{#1}\def\tempb{#2}%
  \ifx\tempa\tempb\relax\href@Urllike{#1}{#2}\else
  \href@Orig{#1}{#2}\fi}
\def\href#1#2{%
  \IfBeginWith{#1}{https://doi.org}%
  {\href@Urllike{#1}{#2}}{\href@Notdoi{#1}{#2}}}
\makeatother

\newlength{\cslhangindent}
\newlength{\csllabelwidth}
\newenvironment{CSLReferences}[3] % #1 hanging-ident, #2 entry spacing
 {% don't indent paragraphs
  \setlength{\parindent}{0pt}
  \ifodd #1 \everypar{\setlength{\hangindent}{\cslhangindent}}\ignorespaces\fi
  \ifnum #2 > 0
  \setlength{\parskip}{#2\baselineskip}
  \fi
 }%
 {}
\usepackage{calc}

\usepackage[top=3.5cm, bottom=3cm, right=1.5cm, left=1.0cm,
            headheight=2.2cm, reversemp, includemp, marginparwidth=4.5cm]{geometry}

\titleformat{\section}
  {\normalfont\sffamily\Large\bfseries}
  {}{0pt}{}
\titleformat{\subsection}
  {\normalfont\sffamily\large\bfseries}
  {}{0pt}{}
\titleformat{\subsubsection}
  {\normalfont\sffamily\bfseries}
  {}{0pt}{}
\titleformat*{\paragraph}
  {\sffamily\normalsize}

\usepackage{fancyhdr}
\makeatletter
\let\ps@plain\ps@fancy
\fancyheadoffset[L]{4.5cm}
\fancyfootoffset[L]{4.5cm}

\definecolor{linky}{rgb}{0.0, 0.5, 1.0}

\newtcolorbox{repobox}
   {colback=red, colframe=red!75!black,
     boxrule=0.5pt, arc=2pt, left=6pt, right=6pt, top=3pt, bottom=3pt}

\newcommand{\ExternalLink}{%
   \tikz[x=1.2ex, y=1.2ex, baseline=-0.05ex]{%
       \begin{scope}[x=1ex, y=1ex]
           \clip (-0.1,-0.1)
               --++ (-0, 1.2)
               --++ (0.6, 0)
               --++ (0, -0.6)
               --++ (0.6, 0)
               --++ (0, -1);
           \path[draw,
               line width = 0.5,
               rounded corners=0.5]
               (0,0) rectangle (1,1);
       \end{scope}
       \path[draw, line width = 0.5] (0.5, 0.5)
           -- (1, 1);
       \path[draw, line width = 0.5] (0.6, 1)
           -- (1, 1) -- (1, 0.6);
       }
   }

\patchcmd{\@maketitle}{center}{flushleft}{}{}
\patchcmd{\@maketitle}{center}{flushleft}{}{}
\patchcmd{\@maketitle}{\LARGE}{\LARGE\sffamily}{}{}
\def\maketitle{{%
  
  \AB@maketitle}}
\makeatletter
\renewcommand\AB@affilsepx{ \protect\Affilfont}
\renewcommand\AB@affilnote[1]{{\bfseries #1}\hspace{3pt}}
\renewcommand{\affil}[2][]%
   {\newaffiltrue\let\AB@blk@and\AB@pand
      \if\relax#1\relax\def\AB@note{\AB@thenote}\else\def\AB@note{#1}%
        \setcounter{Maxaffil}{0}\fi
        \begingroup
        \let\href=\href@Orig
        \let\texttt=\textttOrig
        \let\protect\@unexpandable@protect
        \def\thanks{\protect\thanks}\def\footnote{\protect\footnote}%
        \@temptokena=\expandafter{\AB@authors}%
        {\def\\{\protect\\\protect\Affilfont}\xdef\AB@temp{#2}}%
         \xdef\AB@authors{\the\@temptokena\AB@las\AB@au@str
         \protect\\[\affilsep]\protect\Affilfont\AB@temp}%
         \gdef\AB@las{}\gdef\AB@au@str{}%
        {\def\\{, \ignorespaces}\xdef\AB@temp{#2}}%
        \@temptokena=\expandafter{\AB@affillist}%
        \xdef\AB@affillist{\the\@temptokena \AB@affilsep
          \AB@affilnote{\AB@note}\protect\Affilfont\AB@temp}%
      \endgroup
       \let\AB@affilsep\AB@affilsepx
}
\makeatother

\renewcommand\Affilfont{\sffamily\small\mdseries}
\ifnum 0\ifxetex 1\fi\ifluatex 1\fi=0 % if pdftex
  \usepackage[T1]{fontenc}
  \usepackage[utf8]{inputenc}

\else % if luatex or xelatex
  \ifxetex
    \usepackage{mathspec}
    \usepackage{fontspec}

  \else
    \usepackage{fontspec}
  \fi
  \defaultfontfeatures{Ligatures=TeX,Scale=MatchLowercase}

\fi
\IfFileExists{upquote.sty}{\usepackage{upquote}}{}
\IfFileExists{microtype.sty}{%
\usepackage{microtype}
\UseMicrotypeSet[protrusion]{basicmath} % disable protrusion for tt fonts
}{}

\usepackage{hyperref}
\hypersetup{unicode=true,
            pdftitle={popsynth: A generic astrophysical population synthesis framework},
            pdfborder={0 0 0},
            breaklinks=true}
\let\addcontentslineOrig=\addcontentsline
\def\addcontentsline#1#2#3{\bgroup
  \let\texttt=\textttOrig\addcontentslineOrig{#1}{#2}{#3}\egroup}
\let\markbothOrig\markboth
\def\markboth#1#2{\bgroup
  \let\texttt=\textttOrig\markbothOrig{#1}{#2}\egroup}
\let\markrightOrig\markright
\def\markright#1{\bgroup
  \let\texttt=\textttOrig\markrightOrig{#1}\egroup}

\IfFileExists{parskip.sty}{%
\usepackage{parskip}
}{% else
\setlength{\parindent}{0pt}
\setlength{\parskip}{6pt plus 2pt minus 1pt}
}
\ifx\paragraph\undefined\else
\let\oldparagraph\paragraph
\renewcommand{\paragraph}[1]{\oldparagraph{#1}\mbox{}}
\fi
\ifx\subparagraph\undefined\else
\let\oldsubparagraph\subparagraph
\renewcommand{\subparagraph}[1]{\oldsubparagraph{#1}\mbox{}}
\fi

\title{popsynth: A generic astrophysical population synthesis framework}

        \author[1]{J. Michael Burgess}
          \author[2]{Francesca Capel}
    
      \affil[1]{Max Planck Institute for Extraterrestrial Physics,
Giessenbachstrasse, 85748 Garching, Germany}
      \affil[2]{Technical University of Munich, Boltzmannstrasse 2,
85748 Garching, Germany}
  \date{\vspace{-7ex}}

\begin{document}
\maketitle

\marginpar{

  \begin{flushleft}
  %\hrule
  \sffamily\small

  {\bfseries DOI:} \href{https://doi.org/10.21105/joss.03257}{\color{linky}{10.21105/joss.03257}}

  \vspace{2mm}

  {\bfseries Software}
  \begin{itemize}
    \setlength\itemsep{0em}
    \item \href{https://github.com/openjournals/joss-reviews/issues/3257}{\color{linky}{Review}} \ExternalLink
    \item \href{https://github.com/grburgess/popsynth}{\color{linky}{Repository}} \ExternalLink
    \item \href{https://doi.org/10.5281/zenodo.5109590}{\color{linky}{Archive}} \ExternalLink
  \end{itemize}

  \vspace{2mm}

  \par\noindent\hrulefill\par

  \vspace{2mm}

  {\bfseries Editor:} \href{https://juanjobazan.com}{Juanjo Bazan} \ExternalLink \\
  \vspace{1mm}
    {\bfseries Reviewers:}
  \begin{itemize}
  \setlength\itemsep{0em}
\item \href{https://github.com/HeloiseS}{@HeloiseS}
  \item \href{https://github.com/warrickball}{@warrickball}
    \end{itemize}
    \vspace{2mm}

  {\bfseries Submitted:} 09 April 2021\\
  {\bfseries Published:} 17 July 2021

  \vspace{2mm}
  {\bfseries License}\\
  Authors of papers retain copyright and release the work under a Creative Commons Attribution 4.0 International License (\href{http://creativecommons.org/licenses/by/4.0/}{\color{linky}{CC BY 4.0}}).

  \end{flushleft}
}

\hypertarget{summary}{%
\section{Summary}\label{summary}}

Simulating a survey of fluxes and redshifts (distances) from an
astrophysical population is a routine task. \texttt{popsynth} provides a
generic, object-oriented framework to produce synthetic surveys from
various distributions and luminosity functions, apply selection
functions to the observed variables and store them in a portable (HDF5)
format. Population synthesis routines can be constructed either using
classes or from a serializable YAML format allowing flexibility and
portability. Users can not only sample the luminosity and distance of
the populations, but they can create auxiliary distributions for
parameters which can have arbitrarily complex dependencies on one
another. Thus, users can simulate complex astrophysical populations
which can be used to calibrate analysis frameworks or quickly test
ideas.

\hypertarget{statement-of-need}{%
\section{Statement of need}\label{statement-of-need}}

\texttt{popsynth} provides a generic framework for simulating
astrophysical populations with an easily extensible class inheritance
scheme that allows users to adapt the code to their own needs. As
understanding the rate functions of astrophysical populations (e.g.,
gravitational wave sources, gamma-ray bursts, active galactic nuclei)
becomes an increasingly important field (Loredo \& Hendry, 2019),
researchers develop various ways to estimate these populations from real
data. \texttt{popsynth} provides a way to calibrate these analysis
techniques by producing synthetic data where the inputs are known (e.g.
Mortlock et al., 2019). Moreover, selection effects are an important
part of population analysis and the ability to include this property
when generating a population is vital to the calibration of any survey
analysis method which operates on an incomplete sample.

Similar frameworks exist for simulating data from specific catalogs such
as \texttt{SkyPy} (Amara \& al., 2021) and \texttt{firesong} (Tung et
al., 2021), however, these have much more focused applications and do
not include the ability to impose selection functions.

\hypertarget{procedure}{%
\section{Procedure}\label{procedure}}

Once a rate function and all associated distributions are specified in
\texttt{popsynth}, a numeric integral over the rate function produces
the total rate of objects in the populations. A survey is created by
making a draw from a Poisson distribution with mean equal to the total
rate of objects multiplied by survey duration for the number of objects
in the survey. For each object, the properties such as distance and
luminosity are sampled from their associated distributions. Selection
functions are then applied to latent or observed variables as specified
by the user. Finally, all population objects and variables are returned
in an object that can be serialized to disk for later examination.
Further details on the mathematics, procedure, and details on
customization can be found in the extensive
\href{https://popsynth.readthedocs.io/}{documentation}.

\hypertarget{acknowledgments}{%
\section{Acknowledgments}\label{acknowledgments}}

This project was inspired by conversations with Daniel J. Mortlock
wherein we tried to calibrate an analysis method we will eventually get
around to finishing. Inspiration also came from wanting to generalize
the examples from Will Farr's lecture note (Farr, 2019). J. Michael
Burgess acknowledges support from the Alexander von Humboldt Stiftung.
Francesca Capel acknowledges financial support from the Excellence
Cluster ORIGINS, which is funded by the Deutsche Forschungsgemeinschaft
(DFG, German Research Foundation) under Germany's Excellence Strategy -
EXC-2094-390783311.

\hypertarget{references}{%
\section*{References}\label{references}}
\addcontentsline{toc}{section}{References}

\hypertarget{refs}{}
\begin{CSLReferences}{1}{0}
\leavevmode\hypertarget{ref-skypy}{}%
Amara, A., \& al., et. (2021). SkyPy: A package for modelling the
universe. In \emph{GitHub repository}. GitHub.
\url{https://github.com/skypyproject/skypy}

\leavevmode\hypertarget{ref-selection}{}%
Farr, W. (2019). An example of treating selection effects via summing
over non-detections in stan. In \emph{GitHub repository}. GitHub.
\url{https://github.com/farr/SelectionExample}

\leavevmode\hypertarget{ref-Loredo:2019}{}%
Loredo, T. J., \& Hendry, M. A. (2019). {Multilevel and hierarchical
Bayesian modeling of cosmic populations}. \emph{arXiv},
\emph{astro-ph.IM}. \href{https://arXiv.org}{arXiv.org}

\leavevmode\hypertarget{ref-Mortlock:2019}{}%
Mortlock, D. J., Feeney, S. M., Peiris, H. V., Williamson, A. R., \&
Nissanke, S. M. (2019). {Unbiased Hubble constant estimation from binary
neutron star mergers}. \emph{Physical Review D}, \emph{100}(10), 103523.
\url{https://doi.org/10.1103/physrevd.100.103523}

\leavevmode\hypertarget{ref-firesong}{}%
Tung, C. F., Glauch, T., Larson, M., Pizzuto, A., Reimann, R., \&
Taboada, I. (2021). FIRESONG: A python package to simulate populations
of extragalactic neutrino sources. \emph{Journal of Open Source
Software}, \emph{6}(61), 3194. \url{https://doi.org/10.21105/joss.03194}

\end{CSLReferences}

\end{document}